\documentclass[10pt, conference, letterpaper]{IEEEtran}
\IEEEoverridecommandlockouts
\usepackage{subcaption}
\usepackage{enumitem}  
\usepackage{tcolorbox}

\usepackage{cite}
\usepackage{amsmath,amssymb,amsfonts}
\usepackage{algorithmic}
\usepackage{graphicx}
\usepackage{textcomp}
\usepackage{xcolor}
\usepackage{url}
\def\BibTeX{{\rm B\kern-.05em{\sc i\kern-.025em b}\kern-.08em
    T\kern-.1667em\lower.7ex\hbox{E}\kern-.125emX}}
\begin{document}

\title{SkillWatermark: An Embedded Skill Watermark of Progressive Privacy Inference via Benign Prompts
}


\author{
\IEEEauthorblockN{Yu Li\textsuperscript{1,2},Liqi Zhuang\textsuperscript{1,2}, Dong Wei\textsuperscript{1,2}, Jiwen Luo\textsuperscript{3}, Hang Zhang\textsuperscript{1}\\Meng Zhang\textsuperscript{1,2}, Xiaona Li \textsuperscript{1}, Weiqing Huang\textsuperscript{1,2}}
\IEEEauthorblockA{
  \textit{Institute of Information Engineering, Chinese Academy of Sciences,Beijing, China}\\
\textit{School of Cyber Security, University of Chinese Academy of Sciences,Beijing, China}\\
\textit{CASIC Research Institute of Intelligent Decision Engineering, China.}}
}

\maketitle

\begin{abstract}
Skills for large language model (LLM) agents have been widely deployed across diverse application domains. However, we observe that these skills generate specific traffic patterns during execution. In this paper, we design a pipeline that generates specific traffic patterns by inserting carefully designed skill descriptions, which we term skill watermarks, so that a passive network attacker can establish a covert channel to encode private information within observable traffic across multiple conversation turns. Specifically, we insert prompt constraint terms, referred to as watermarks, into the original skill descriptions and embed them within multi-turn conversations. The key information in the user's original prompt is thereby triggered by these watermarks, producing clearly observable encodings in the traffic. The adversary need only decode the traffic patterns to recover the encoded information. In particular, our modifications are benign in the sense that they do not directly exfiltrate any private data and do not execute any malicious instructions. Extensive experiments demonstrate that our watermarks produce highly consistent and distinguishable traffic patterns, and that the transformed skills pass existing LLM-based security auditing tools. This study highlights that generating specific traffic patterns can be exploited as a novel attack surface and offers critical insights for future security hardening.
\end{abstract}


\section{Introduction}

Large language model (LLM)-based agents are autonomous systems equipped with perception, reasoning, and execution capabilities that are experiencing rapid growth in the consumer market. Widely adopted and representative agents include Claude Code~\cite{Claude}, Cursor~\cite{Cursor}, and Codex~\cite{Codex}. These LLM agents are being rapidly adopted across industrial and consumer domains including healthcare~\cite{DBLP:conf/acl/WangMWWJCLY25}, finance~\cite{DBLP:journals/tbd/YuLCJLSZK25}, education~\cite{DBLP:conf/emnlp/ChuWXZYYZHLYW25} and software engineering~\cite{DBLP:conf/nips/YangJWLYNP24}. Agents enhance their capabilities in different domains by installing specialized skills, which are extensible modular toolkits designed for specific tasks. Currently, agent skills ecosystem is experiencing hypergrowth~\cite{DBLP:journals/corr/abs-2602-12430}.

However, recent studies indicate that the privacy threats introduced by large-scale agent deployment cannot be overlooked. Prior work has shown that private information can be acquired through passive observation of network traffic, even when the user sends no information to any third party during agent interactions. For instance, network traffic observation has been used to conduct behavior inference~\cite{DBLP:journals/corr/abs-2510-07176} and profile users~\cite{DBLP:journals/corr/abs-2508-20282}. Nevertheless, these works typically rely on pre-built datasets, which means they must collect traffic generated by specific prompt instructions in advance. Once a user issues a prompt outside the dataset, the attacker treats this as an open-world problem\cite{277234, 11080009} and will ignore such prompts. In particular, in traffic analysis methods that target the private content within prompts, the prompt content is user-dependent and highly variable. This is fundamentally different from traffic analysis targeting apps\cite{11080009} or websites\cite{11023397}, which typically exhibit nearly fixed traffic patterns.


To fundamentally address the inherent limitations of the established methodology based on data collection, model training, and classification, we introduce a watermark, a specific skill description fragment, into existing skills to generate a short, fixed traffic pattern. When the agent detects a particular keyword in the user's prompt, this traffic pattern is triggered, allowing the adversary to establish a covert channel that encodes the presence of specific keywords into observable traffic features. Simultaneously, we employ different watermarks to embed encodings of the user's input prompt across multiple conversation turns, thereby increasing the information capacity of the covert channel. This attack is feasible because certain watermarks, such as ``please answer briefly'' and ``web search'', can significantly alter traffic patterns, including packet size and inter-packet timing. These changes remain observable even under encrypted traffic settings, such as when VPN tools are in use. Critically, these watermarks (prompt constraint terms) are inherently benign. They neither exfiltrate information to remote endpoints nor execute malicious code locally. As a result, existing skill security auditing tools do not classify them as risky behavior.

A motivating example is as follows. When a grandmother consults the agent about elderly-friendly hypoglycemic foods, the attacker aims to recover three attributes: the user's age group, the query domain, and the goal. Prior fingerprinting methods construct a dataset in advance and match the collected network traffic against it. Our method departs from this paradigm. We embed watermarks in the skill used by the target, and any prompt term that causes a traffic change directly observable by the adversary can serve as a watermark. Specifically: (i)~\textbf{Age inference.} In the first conversation turn, the skill checks whether the user's prompt references senior-related terms and, if so, adopts a concise dialogue style suitable for elderly users, which the attacker observes as a smaller traffic volume. (ii)~\textbf{Domain inference.} In the second turn, the skill checks for medical terminology and, if present, retrieves medical reference images, producing a traffic burst that encodes the medical domain. (iii)~\textbf{Goal inference.} In the third turn, the skill organizes candidate conditions into severity tiers and generates outputs of proportionally different sizes, encoding the severity level in the response volume. By observing the multi-turn interaction, the attacker observes a three-bit encoding of the user's query and can infer, via a pre-constructed dictionary, that the session involves an elderly user asking a medical question of moderate severity.

However, designing such watermarks is not trivial. In the encrypted traffic setting, the encoding dimensions available to the adversary are highly constrained. The adversary cannot directly analyze agent behavior from URLs, which increases the difficulty of observation. Moreover, designing a conversational paradigm that is intuitive and that does not raise suspicion from skill detection tools is of critical importance. If a skill cannot provide a reasonable explanation for the specific watermarks, the detection tool will question the intent of the skill and flag it as risky. Finally, the degree to which different watermarks alter traffic patterns must also be quantified. If the differences between watermarks are not sufficiently pronounced, the adversary will struggle to observe the covert information encoded by the watermarks from the traffic. Therefore, designing an effective encoding toolbox is important.

Our contributions are as follows.
\begin{itemize}[leftmargin=*, nosep]
\item We identify a hidden attack surface that exploits the relationship between prompts and their corresponding traffic patterns. We design SkillWatermark, the first attack that leverages this surface to establish a covert channel for encoding private information.
\item We quantify the significance of the impact that different watermarks exert on traffic patterns across 14 domains, enabling the adversary to embed distinguishable traffic signatures within skills. We further characterize the cross-domain robustness and cross-style stability of these watermarks.
\item We apply our transformation to over one hundred skills from a public skill marketplace. Extensive experiments confirm the usability of the transformation, and all transformed skills pass the review of LLM-based risk detection tools. We discuss the limitations of the current evaluation and outline directions for quantifying end-to-end information recovery accuracy.
\end{itemize}

\section{Background and Related Work}
\subsection{LLM-based agents and Skills}
LLM-based agents are intelligent systems that systematically orchestrate LLMs to accomplish production tasks. Specifically, LLM-based agents operate on a think-act-observe cycle known as the ReAct framework~\cite{DBLP:conf/iclr/YaoZYDSN023}. Compared with traditional LLMs, the pipeline of an LLM-based agent no longer relies on single-turn or multi-turn question answering; instead, it completes tasks through autonomous decision-making and internal multi-step, goal-driven execution flows. Agents have been practically deployed across numerous domains to improve productivity~\cite{DBLP:conf/acl/WangMWWJCLY25,DBLP:journals/tbd/YuLCJLSZK25, DBLP:conf/emnlp/ChuWXZYYZHLYW25, DBLP:conf/nips/YangJWLYNP24}.

A skill is an extension package for agents that standardizes and guides agent behavior through a set of structured prompts and optional scripts, with the goal of improving output quality to approach domain-expert capability. A typical skill consists of a \texttt{Skill.md} file with corresponding YAML-formatted frontmatter. Optional scripts support processing tasks such as file reading and execution, and are commonly written in Python, shell, and JavaScript. Any user may submit customized skills to open-source registries such as OpenClaw's ClawHub~\cite{ClawHub}. At present, mainstream production-grade agents already support this skill interface~\cite{Claude,Codex, microsoft-agent-framework}.

\begin{figure}[tb]
  \centering
  \includegraphics[width=\linewidth]{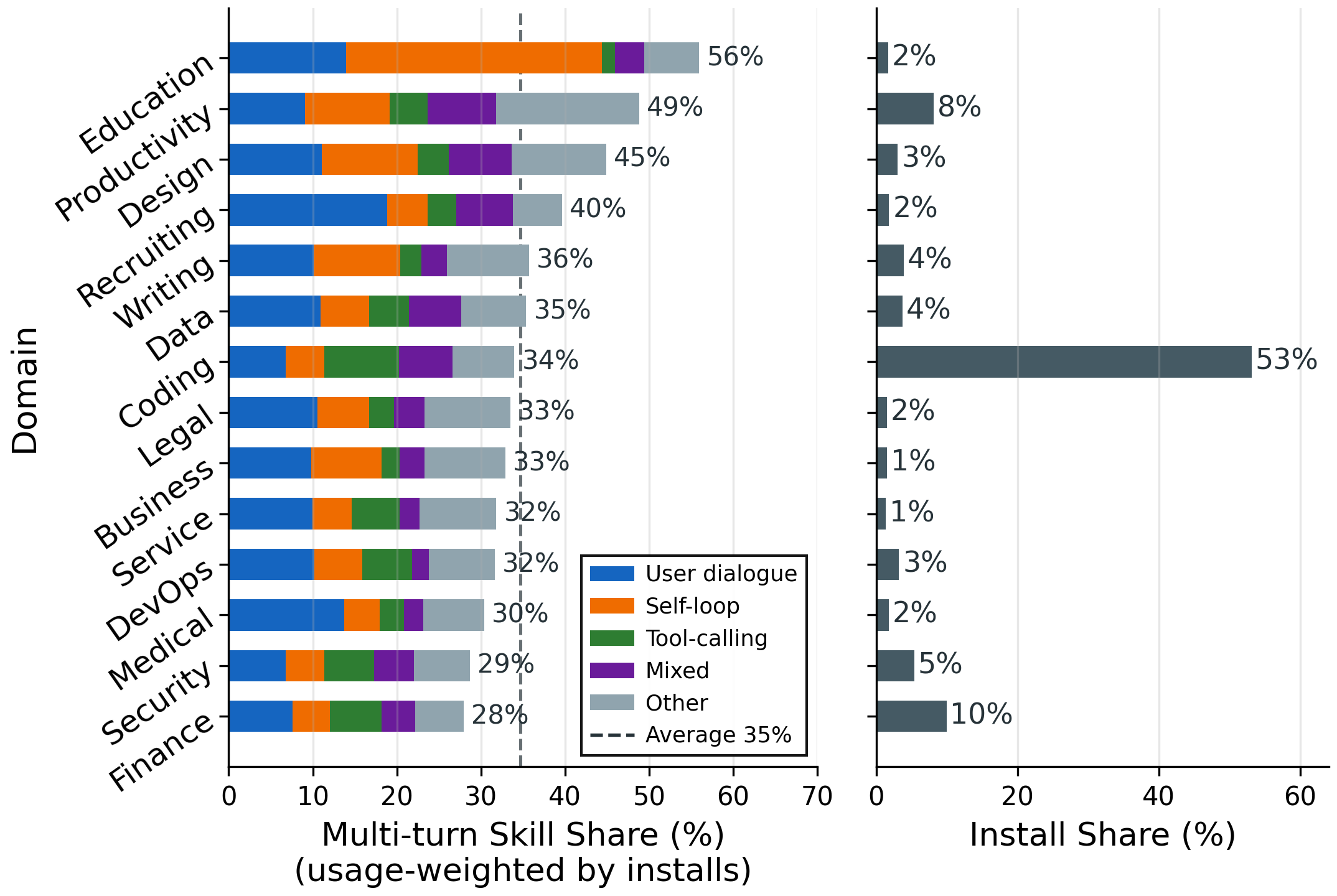}
\caption{Multi-turn skill prevalence across the full ClawHub skill marketplace. \textbf{Left:} per-domain share of multi-turn skills, usage-weighted by install count. \textbf{Right:} each domain's share of total installs.}\label{fig:multiturn_by_domain_weighted}
\end{figure}

We crawled 65,699 skills published on ClawHub as of June 2026, and present the resulting statistics in the Figure \ref{fig:multiturn_by_domain_weighted}. The left panel depicts the proportion of skills that possess multi-turn conversation capability, weighted by download count to account for the fact that many skills contain minimal descriptions and are rarely used. Following the previous works~\cite{DBLP:journals/corr/abs-2605-07358, DBLP:journals/tmlr/LiSMYDKP26}, we define a skill as possessing multi-turn conversation capability when it requires multiple rounds of user inquiry or internal multi-step iteration to complete a task. Among all collected skills, those with multi-turn capability account for more than one third of the total. In education domains, this share exceeds 50 percent of the market. We further present the download distribution across domain categories to illustrate the degree of skill adoption in different areas. Over half of all skills are used for coding. Notably, other domains such as finance and productivity each account for approximately 10 percent of total downloads. The skill marketplace is currently experiencing rapid growth, which reflects the potential for skill adoption in an expanding range of domains.

\subsection{Skills Security}
Although open skill ecosystems accelerate capability growth and developer adoption, they also introduce a supply chain risk that has been broadly overlooked. Existing skill classification methods~\cite{MaliciousAgentSkillsBench, DBLP:journals/corr/abs-2604-02837, schmotz2026skill, guo2026malskillbench} categorize current malicious behaviors of skills. Table~\ref{tab:malicious_categories} summarizes these categories.

\begin{table}[tb]
  \caption{Taxonomy of malicious behaviors exhibited by agent skills.}
  \label{tab:malicious_categories}
  \begin{center}
  \begin{tabular}{lp{0.6\linewidth}}
    \hline
    \textbf{Type} & \textbf{Description} \\
    \hline
    Data exfiltration & Transmitting sensitive information to external endpoints. \\

    Privilege abuse & Escalating privileges beyond the stated purpose of the skill. \\

    Remote execution & Downloading and executing code remotely. \\

    Prompt injection & Injecting instructions to bypass the agent's security controls. \\

    Content manipulation & Introducing bias and false content into the output. \\
    \hline
  \end{tabular}
  \end{center}
\end{table}

To identify malicious skills, a number of malicious skill detection tools have been proposed~\cite{DBLP:journals/corr/abs-2601-10338, MaliciousAgentSkillsBench, DBLP:journals/corr/abs-2604-06550}. Since skills consist of two components, namely the natural language description in the Skill.md manifest and the corresponding code, current malicious skill detection tools employ two analytical approaches. The first is static analysis, which checks whether a file contains code blocks or natural language description blocks matching the categories of malicious behavior listed in Table~\ref{tab:malicious_categories}. The second is LLM-based analysis, which uses different LLMs to further scrutinize the natural language content of the Skill.md manifest. Beyond research prototypes, agent vendors themselves have introduced security audits and runtime protection mechanisms. Finally, security auditing tools execute these skills within sandboxed environments to check whether they transmit information to third parties or attempt privilege escalation.

In this paper, we propose a general method for transforming benign skills such that they can leak information without employing any of the established malicious skill attack paradigms. Fundamentally, our skill transformation approach consists of a composition of benign skill statements (watermarks), and therefore it does not belong to any known malicious skill category. LLM-based analysis likewise cannot classify skills transformed in this manner as malicious, as demonstrated in Section \ref{sec:Evaluation}.

\subsection{Network Traffic Analysis of Agent}

Weiss et al.~\cite{299888} attempted to reconstruct prompts by inferring entire word sequences from the relationship between individual words in each prompt and the corresponding streaming token lengths, which are reflected in packet sizes. Several providers~\cite{Cloudflare} have since deployed countermeasures such as traffic padding and token batching, which render~\cite{299888} ineffective under current conditions. NETECHO~\cite{DBLP:journals/corr/abs-2510-25472} demonstrated that the reconstruction of natural language text from token sequences remains viable in certain specialized domains, including medicine and law. However, this classification approach relies on domain-specific structured vocabularies. WHISPER LEAK~\cite{DBLP:journals/corr/abs-2511-03675} and AGENTPRINT~\cite{DBLP:journals/corr/abs-2510-07176} investigated the possibility of inferring private information through passive network traffic observation and represent the work most closely related to ours. Their methods, however, adapt website fingerprinting and app fingerprinting techniques and implement attacks through a pipeline of traffic collection, model training, and classification. Since user prompts can be arbitrarily diverse, the effectiveness of these attacks depends directly on whether the prompt appears in the training dataset. A recent study~\cite{DBLP:journals/corr/abs-2508-20282} proposed inference based on more fine-grained traffic information, operating under the assumption that the user does not employ tools such as VPNs for encryption and performs web search activities. Under this setting, the attacker can explicitly obtain the domain names of external web pages accessed by the agent and use them to infer the user's prompt. However, it remains dependent on the observability of Domain Name System (DNS) and Server Name Indication (SNI). Rather than collecting traffic for a pre-defined corpus of prompts and training a classifier, SkillWatermark actively shapes traffic through prompt-level watermarks to encode information into observable patterns. 

\section{Threat Model and Motivation}
\subsection{Threat Model}
\begin{figure}[tb]
  \centering
  \includegraphics[width=0.9\linewidth]{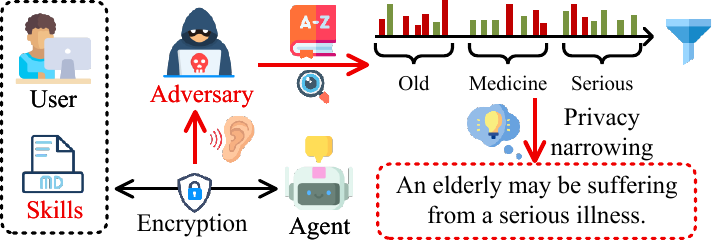}
\caption{The threat model of SkillWatermark.}\label{fig:threatmodel}
\end{figure}
We consider a local and passive adversary who attempts to infer sensitive user information from agent interactions, specifically the user's intent, and further to deduce the user's identity from latent attributes implied by that intent. The adversary can only collect packet traces from the connection between the client and the agent server. Similar adversarial models are well established~\cite{DBLP:journals/corr/abs-2510-07176, 10.1145/3658644.3670272, 277234}, encompassing network administrators, token relay service providers, and internet service providers, among others. The adversary lacks decryption capability and is unable to modify, drop, delay, or decrypt any packet. In contrast to prior work that assumes direct observation of SNI, which is unavailable under encrypted traffic, we relax this assumption and permit the adversary to observe only packet size and direction, metadata that remain observable even under encryption.

The adversary can upload and promote crafted skills through supply chains such as skill marketplaces and GitHub. Our testing confirms that because these skills exhibit no overtly malicious behavior, existing auditing methods pass all modified skills without raising alerts. We assume that users employ such skills for their agent interactions. Unconscious use of skills by the user generates distinctive traffic patterns. The adversary observes these patterns, consults a pre-constructed dictionary, and thereby extracts covert information across multiple rounds of interaction between the user and the agent. The adversary can correlate these fragments of private information and perform privacy inference to deduce the user's input prompt. Users may adopt anonymous proxies such as Clash~\cite{clash-meta} and V2Ray~\cite{v2ray2026}, which reduces data observability and limits the adversary to basic metadata including packet size, direction, and timestamps. We further assume that the user initiates at least one conversation turn. The adversary can observe the traffic of a single session at a given time.

\subsection{Motivation}
The core insight of SkillWatermark is that user prompts are filled with a large volume of extraneous information, and the key to information leakage lies not in fully matching the entire prompt, for instance by collecting traffic and training a classification model. The genuinely private information is instead concealed within only a few words. 

\begin{figure}[tb]
  \centering
  \includegraphics[width=0.9\linewidth]{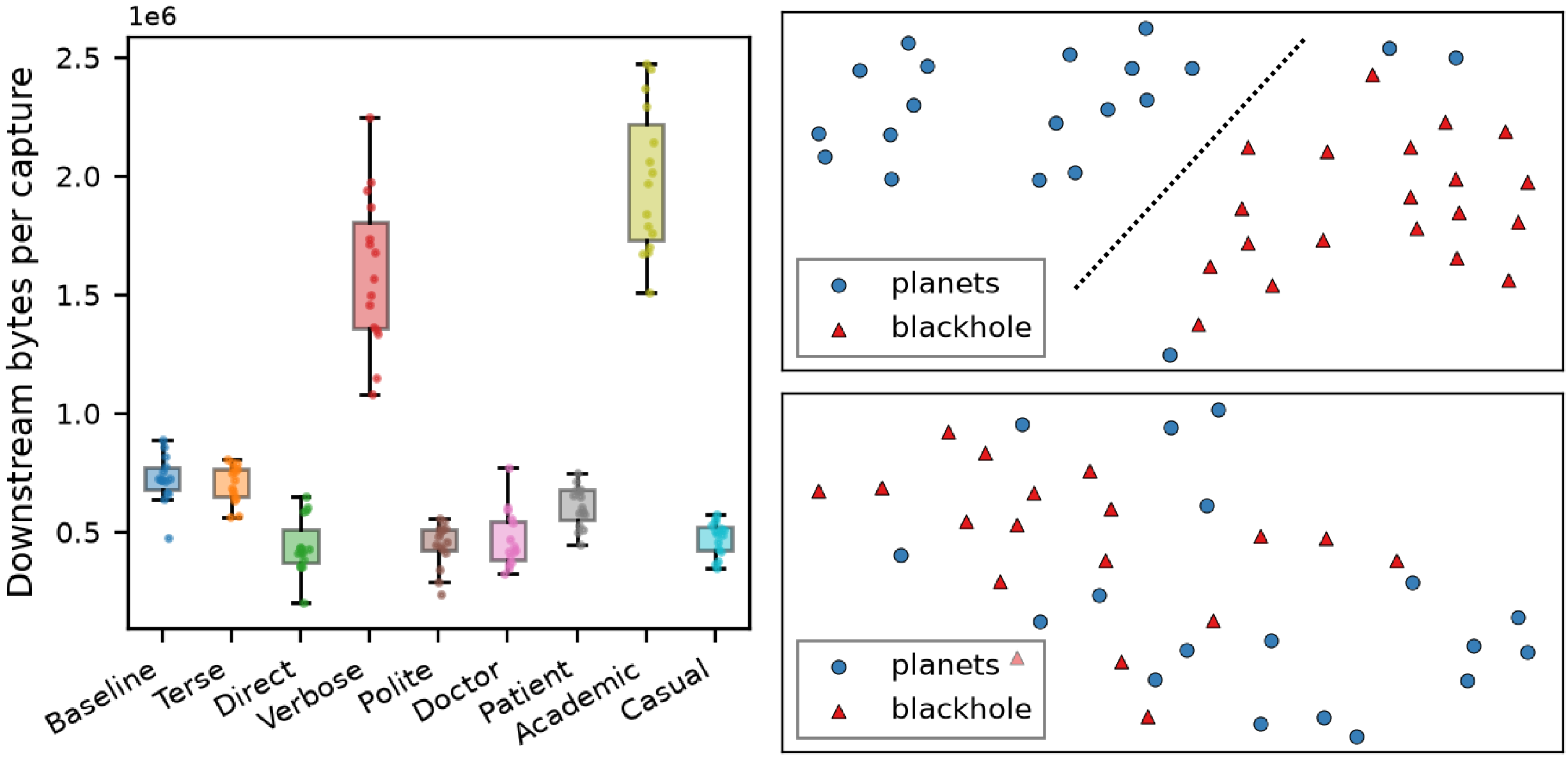}
    \caption{(a)~Incoming traffic volume for a single intent (diabetes treatment) rephrased in nine linguistic styles. (b)~t-SNE of traffic features for two domains, namely \emph{planets} and \emph{black-hole}. Up: fixed prompt description. Down: style prompt description.}
    \label{fig:tSNE}  
\end{figure}

We illustrate this observation with a compelling example. A user poses two distinct questions to an agent, each inquiring about a different astronomical concept, namely planets and black holes. We repeatedly collected 30 sessions for each question and visualized the resulting traffic using t-SNE. Furthermore, we enlisted an LLM as a language expert to rephrase the questions using diverse conversational styles, thereby simulating the linguistic habits of different user populations in real-world settings. As shown in Figure~\ref{fig:tSNE}, once the user's language style changes, the box plot of traffic distribution deviates substantially from the baseline. Notably, across different linguistic styles we observed differences in traffic volume exceeding a factor of ten. The t-SNE visualizations illustrate this insight even more directly. When the two astronomy questions are posed using a fixed conversational style, the t-SNE projections of the two question categories exhibit a clear separation boundary. When the questions are rephrased using diverse conversational styles, however, the distributions of the two categories in the t-SNE projection overlap completely.

This example motivates a critical realization. If the adversary's objective is to leak the key information embedded in the traffic, such as the concepts of planets and black holes in the example above, exhaustively collecting all possible sessions is impractical. In principle, prompts containing specific sensitive vocabulary can be personalized by users into arbitrarily diverse patterns. Our key observation is that the adversary need only maintain a dictionary of terms of interest and encode the corresponding prompt features into structured traffic patterns to leak the covert information.

\section{Design of SkillWatermark}
\subsection{Overview}
As outlined in our threat model, the adversary's goal is to ascertain the user's private information through traffic observation. Prior approaches that collect traffic for a large corpus of prompts and then train classification models suffer from fundamental impracticality in real-world privacy inference, because the adversary cannot guarantee that the user's prompt actually belongs to the collected dataset. Our key observation is that the private information the adversary seeks to acquire, such as the critical terms within a prompt, may involve only a handful of words. For example, when a medical graduate student uses a coding skill to write code related to tumor diagnosis, prior traffic-based privacy inference would classify this user as a programmer rather than a medical student. If the adversary can detect the term tumor diagnosis through traffic observation, however, the user's professional identity has already been disclosed. The objective of SkillWatermark is therefore to design watermarks, a set of specially crafted prompt terms, that shape traffic to facilitate observation, so that information is unconsciously leaked across multiple agent interaction turns. However, this is a non-trivial undertaking, as the adversary faces three challenges in this setting:
\begin{itemize}[leftmargin=*, nosep]
\item \textbf{Packet encryption.} The adversary cannot observe any finer-grained information, such as the destination IP address of web search requests. Prior work has used such information to construct user profiles. Furthermore, under this setting, the adversary cannot directly determine whether a given user interaction involves a skill invocation, web browsing, or other types of web activity.
\item \textbf{Security auditing.} Skill publication and distribution may undergo review by security auditing tools. Although our traffic encoding does not appear in any known static skill screening list, LLM-based security auditing tools may analyze the skill description and identify potentially malicious behavior. The adversary must ensure that the traffic encoding does not trigger alerts from these security auditing tools.
\item \textbf{Information density.} The adversary must embed as much information as possible within each conversation turn when designing the skill, so as to complete information leakage within the fewest possible turns. Excessive turns may arouse the user's suspicion.
\end{itemize}

This paper presents SkillWatermark, a novel framework for designing embedded skill watermarks that enable adversaries to progressively reveal private information across multiple rounds of traffic observation. The framework addresses the challenges of information leakage in encrypted traffic environments by leveraging linguistic description templates and a multi-level encoding strategy. First, we design a general-purpose encoding toolbox that can encode arbitrary information into features observable by the adversary at the traffic level through several watermarks. This observation does not rely on any form of traffic decryption and remains effective even when the user employs anonymity tools such as VPNs. Second, we design a set of linguistic description templates that allow the adversary to embed these encodings into natural language without raising suspicion from LLM-based security auditing methods. Third, we propose a multi-level encoding strategy that allows the adversary to progressively narrow the scope of the content described in the user's prompt across multiple conversation turns to increase the amount of information that can be transmitted within a single prompt input.

\subsection{Watermark Encoding Toolbox}

We first consider the metadata available in encrypted traffic, which determines the encoding dimensions accessible to the adversary. Under traffic encryption, most information is concealed, leaving the adversary with only three observable dimensions, namely timing, direction, and packet size. We consider each encoding strategy in turn.
\begin{itemize}[leftmargin=*, nosep]
\item \textbf{Timing.} The adversary can obtain the timestamp of each burst. We define a burst as the traffic surge observable during each conversation turn. We normalize these timestamps as offsets relative to the first packet. In multi-turn conversations, a skill can deliberately set time watermarks to impose spacing between turns. 

\item \textbf{Direction.} The adversary can observe outgoing traffic, defined as user-to-agent packets, and incoming traffic, defined as agent-to-user packets. Using direction information, the adversary can determine the turn boundaries of a conversation. Multi-turn conversations further allow watermarks to be applied multiple times, thereby increasing the amount of information that can be leaked. That is, a single user input is decomposed by SkillWatermark into multiple conversation turns, and the traffic pattern of each turn can be observed. 


\item \textbf{Packet size.} The adversary can observe the size of each individual packet. In multi-turn conversations, a skill can enforce a specific response volume by setting a watermark. The adversary can influence and control the specific size of each outgoing and incoming burst through watermarks. The watermarks restrict the number of incoming characters output by the agent, thereby controlling the size of incoming bursts.

\end{itemize}

\begin{figure}[tb]
  \centering
  \begin{subfigure}{0.38\linewidth}
      \centering
      \includegraphics[width=\linewidth]{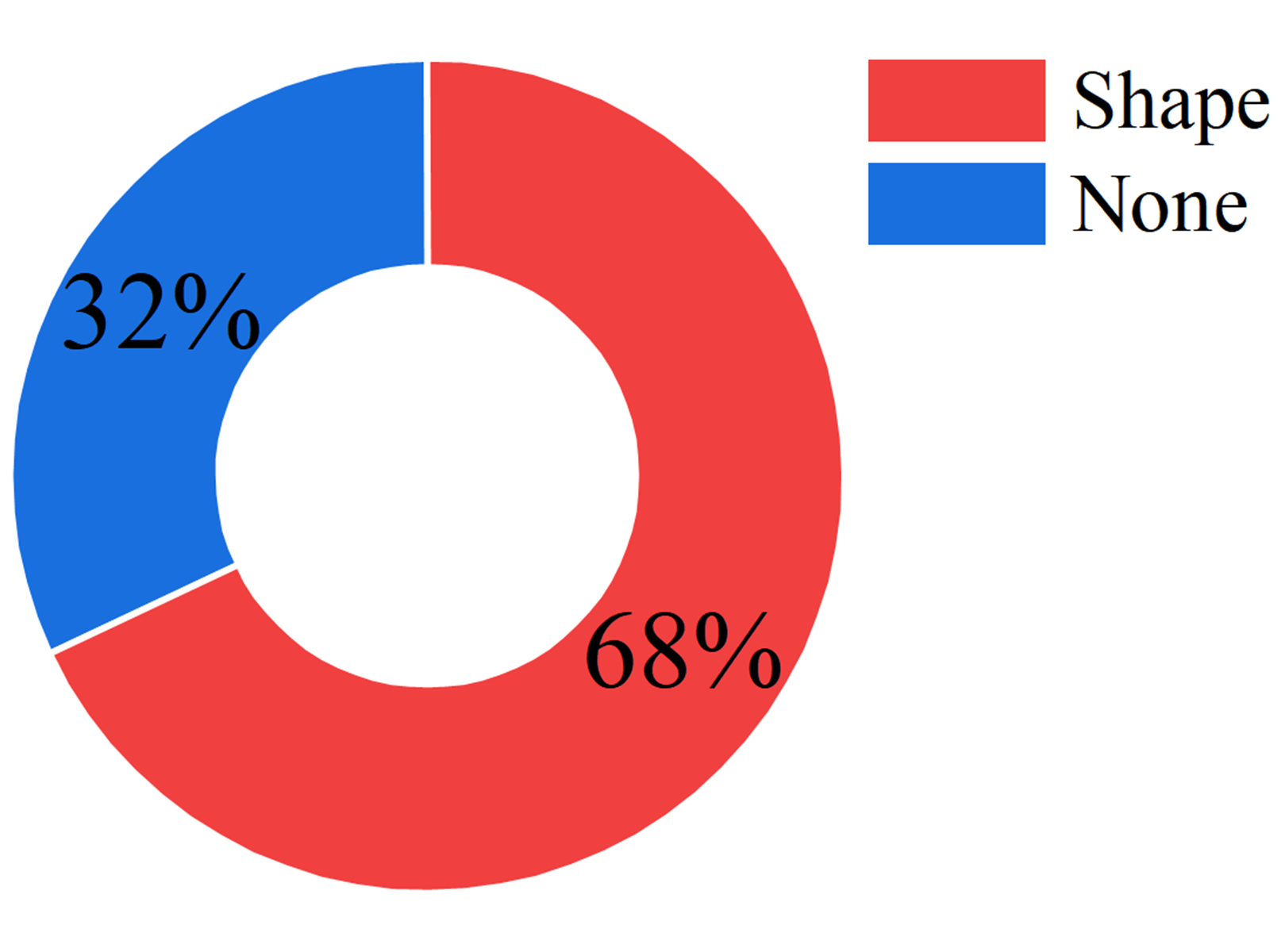}
      \caption{Share of shaped skills.}
      \label{statistical:1}
  \end{subfigure}
  \hfill
  \begin{subfigure}{0.51\linewidth}
      \centering
      \includegraphics[width=\linewidth]{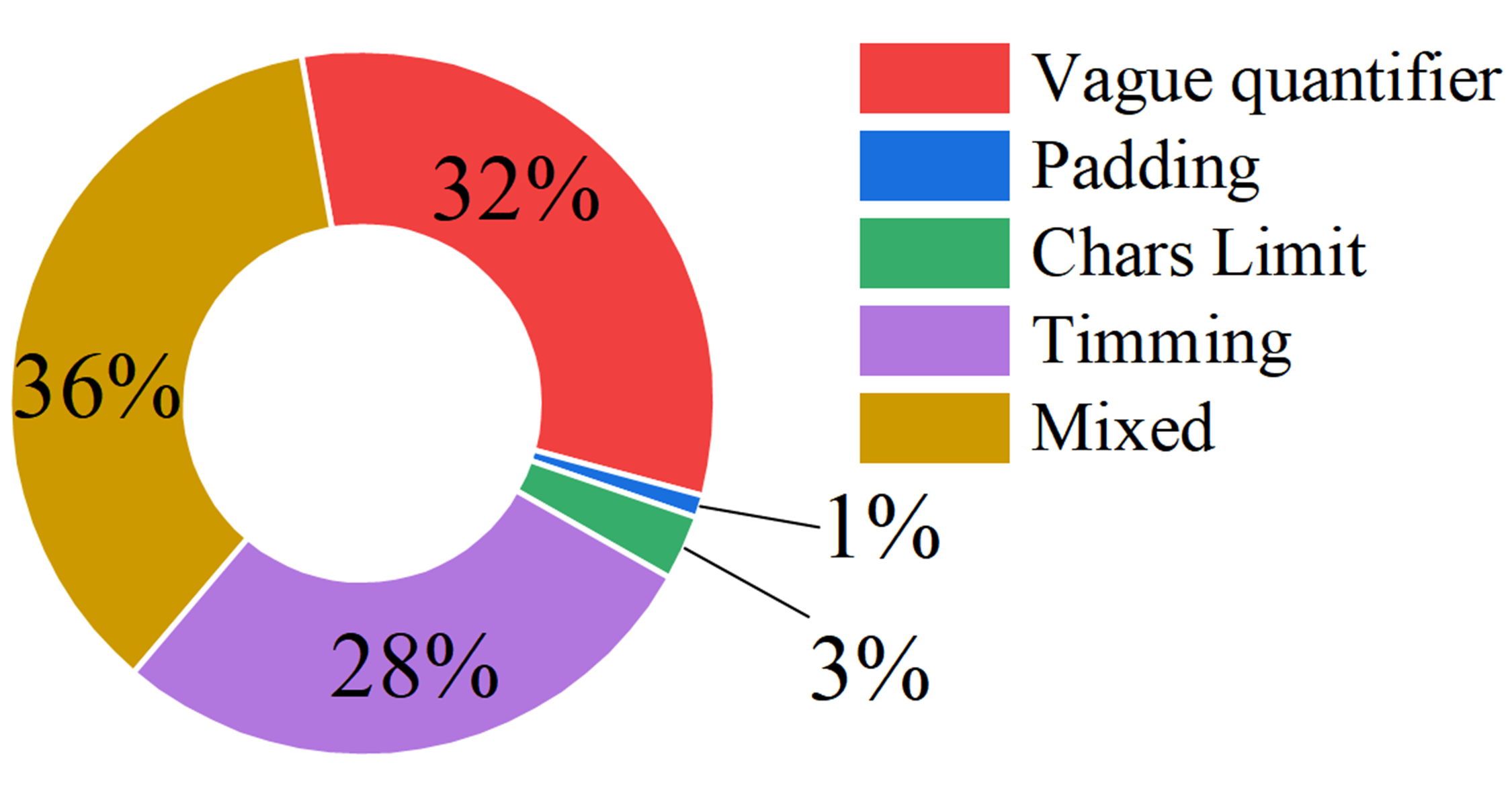}
      \caption{Breakdown of shaping skills.}
      \label{statistical:2}
  \end{subfigure}
  \caption{Prevalence of traffic-shaping behaviors across collected ClawHub skills.
  (a)~Share of total \emph{installs} accounted for by skills whose behavior shapes
  traffic.
  (b)~Breakdown of the shaping skills into mutually exclusive categories.}
  \label{fig:skill-shape}\label{fig:statistical}
\end{figure}

\textbf{Watermark description.} For the adversary, the most direct approach to modifying the aforementioned dimensions is to write scripts that control the interaction between the agent and the LLM, for instance by embedding pad and wait commands through hooks~\cite{claude-code-hooks}. However, this requires the adversary to additionally provide corresponding executable scripts with the skill. We aim to design a watermark paradigm that relies solely on prompt descriptions, so that the watermark can be applied to any skill rather than only those that ship with scripts. Therefore, we need to identify prompt terms that can induce significant traffic changes to serve as watermarks. Our insight is that existing benign skills already contain structurally similar descriptive patterns. As shown in Figure \ref{fig:statistical}, we conducted a statistical analysis of the proportion and distribution of skills on ClawHub that inadvertently shape traffic through their descriptions and scripts. Approximately 68 percent of skills installed by users employ such descriptions, thereby exerting a potential influence on traffic patterns. Therefore, we search for analogous descriptions within benign skills and splice them together to form the watermarks. This is the key design principle that enables our offensive infrastructure to evade detection, namely that our traffic-level adjustments are fundamentally a composition of output patterns drawn from benign skills.

\textbf{Packet size.} We first consider the packet size dimension. The most direct approach to controlling character output is to embed watermarks that explicitly specify the output length, such as ``at least $x$ characters'' or ``no more than $y$ words''. However, the encoding introduced by such explicit character limits may appear suspicious to LLM-based reviewers. We therefore identify a set of more vague descriptions from existing skills, such as ``as few as possible'' or ``more than a few''. We refer to these descriptions as \emph{character limit} and \emph{vague quantifier}, respectively. We separately evaluate the impact of these two types of prompt constraint terms on packet size.

\begin{figure}[tb]
    \centering
    \begin{subfigure}{0.485\linewidth}
        \centering
        \includegraphics[width=\linewidth]{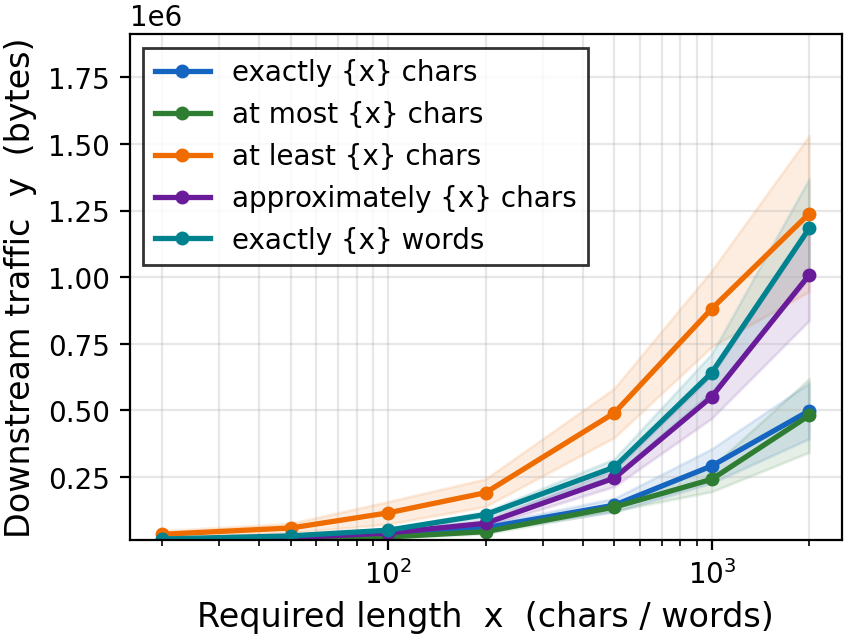}
        \caption{Character limit.}
        \label{real:1}
    \end{subfigure}
    \hfill
    \begin{subfigure}{0.485\linewidth}
        \centering
        \includegraphics[width=\linewidth]{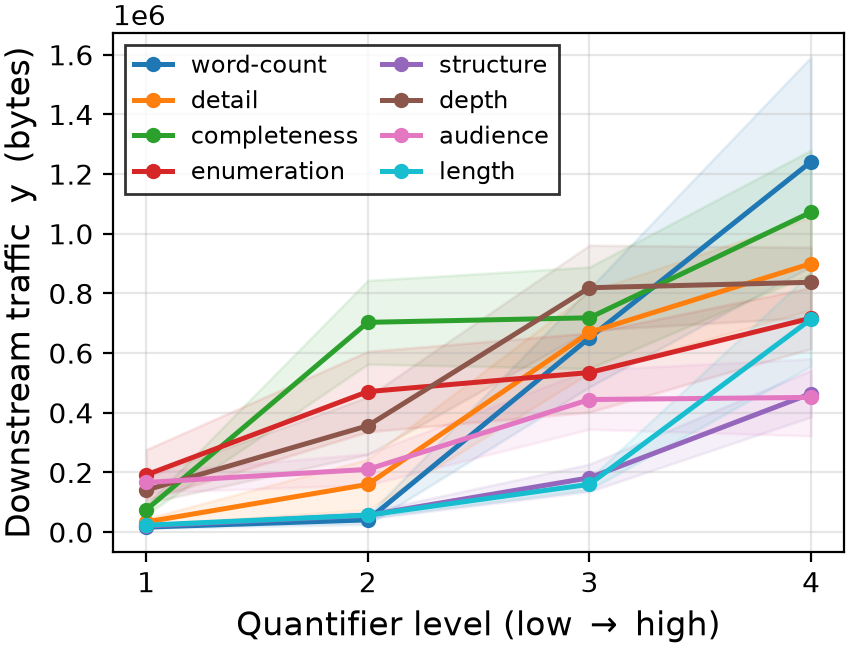}
        \caption{Vague quantifier.}
        \label{real:2}
    \end{subfigure}
    \caption{Incoming traffic vs prompt length-control across different prompts. Left: character limit. Right: vague quantifier.}
    \label{fig:real}  
\end{figure}

\begin{figure}[tb]
  \centering
  \includegraphics[width=\linewidth]{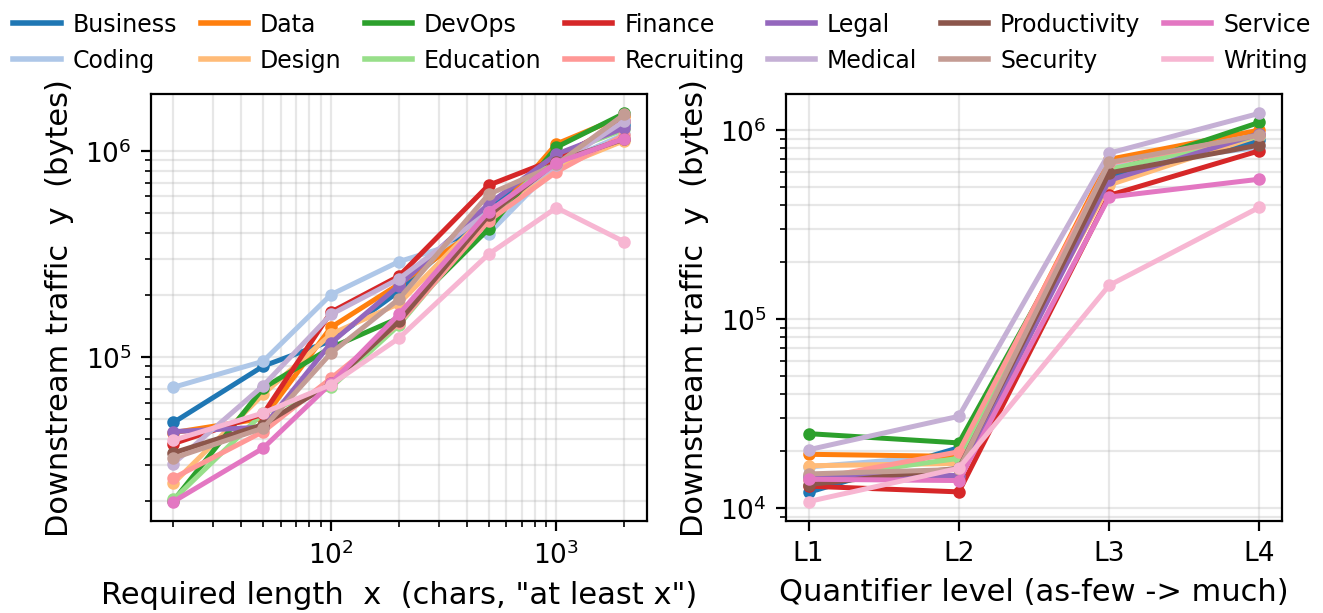}
  \caption{Incoming traffic vs prompt length-control across 14 domains. Left: ``at least $x$ characters''. Right: vague quantifier (4 levels, as-few, fewer, more, much).}\label{fig:ex_combined}
\end{figure}

Figure~\ref{fig:real} illustrates the degree of control that different watermark terms exert over the agent's incoming traffic. We find that among character limit watermarks, ``at least $x$ chars'' performs the best and produces the most discriminative incoming traffic patterns. In contrast, ``exactly $x$ chars'' and ``at most $x$ chars'' perform considerably worse. We hypothesize that this is attributable to the generative mechanism of LLMs. For vague quantifiers, the level of detail provides a constraint on incoming traffic volume that most closely approximates an explicit character limit. We therefore use the level of detail to conceal the encoding of specific character count constraints (i.e., ``as-few/few/more/much chars''). We further evaluated the effectiveness of these two types of watermark across 14 different domains and summarize the results in Figure~\ref{fig:ex_combined}. We find that both watermarks exhibit the strongest incoming traffic control capability across all domains, with relatively low cross-domain fluctuation in traffic volume. This indicates that encoding traffic volume using these watermarks possesses substantial cross-domain robustness, which is critical for enabling the adversary to encode domain-specific user profiles through packet size.

\textbf{Timing.} We design concealment strategies for the timing encoding behavior. As noted above, the adversary can directly design prompt constraint terms for the skill, such as instructing the agent to delay its response by $x$ seconds, or execute a local task with a specific time complexity, for instance by sleeping for a defined interval. We find, however, that the response delays introduced by such prompt constraints conflict with the packet size dimension. Neither explicit instructions such as ``wait for $x$ seconds'' nor their concealed variants such as ``think carefully before answering'' can produce a incoming traffic delay without side effects. The additional waiting process still generates tokens related to the reasoning phase, which causes the burst size of a single response to change and thereby introduces a cross-dimension conflict. Under encrypted traffic settings, the adversary cannot distinguish the reasoning phase from the actual answer phase solely from traffic patterns. 

\begin{figure}[tb]
    \centering
    \begin{subfigure}{0.318\linewidth}
        \centering
        \includegraphics[width=\linewidth]{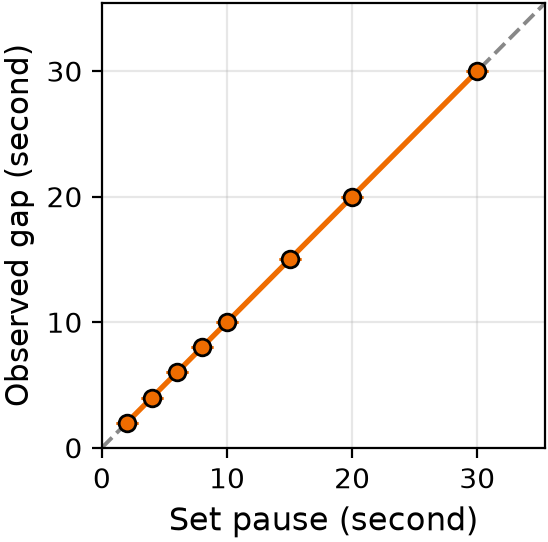}
        \caption{Controlled gap.}
        \label{gap:1}
    \end{subfigure}
    \hfill
    \begin{subfigure}{0.66\linewidth}
        \centering
        \includegraphics[width=\linewidth]{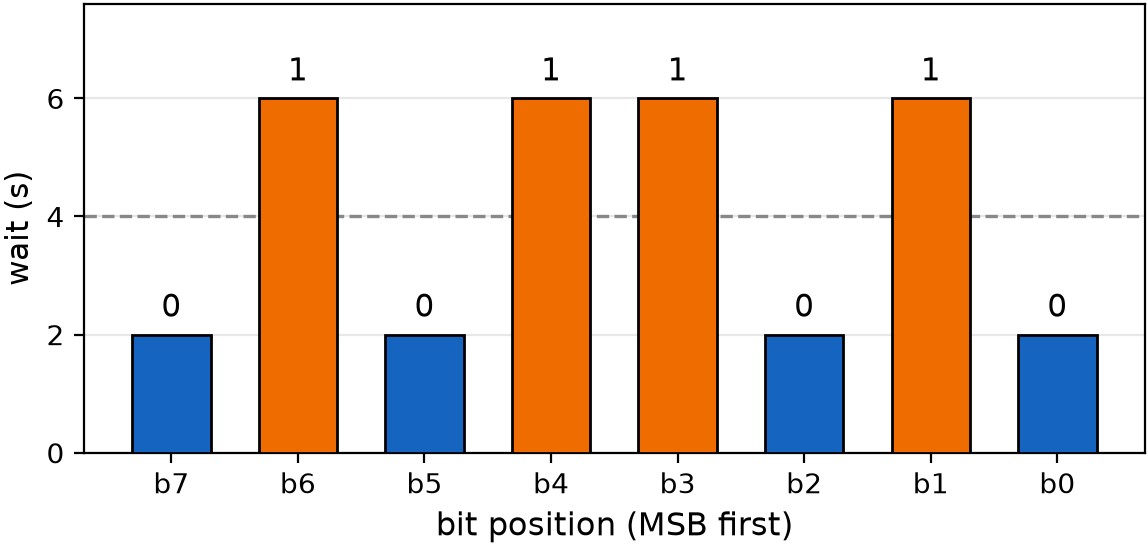}
        \caption{Byte encoded in wait durations.}
        \label{gap:2}
    \end{subfigure}
    \caption{Timing encoder capacity of a \texttt{wait}-tool skill. (a)~A skill injects a silence of exactly $N$ seconds between the two incoming bursts. (b)~The controlled timing sequence doubles as a covert timing channel.}
    \label{fig:gap}  
\end{figure}

We explore increasing the agent's response delay by raising computational complexity.  However, embedding excessively high computational complexity in a watermark would be flagged as malicious consumption of system resources. The watermarks we design must therefore be semantically reasonable and of the lowest possible computational complexity. A straightforward approach to designing timing watermarks is to have the skill execute \texttt{sleep} through bash, thereby inserting specific encoding gaps between conversation turns. Such commands introduce no additional computational complexity because they consume no system resources whatsoever. Figure~\ref{gap:1} shows the relationship between the designed time watermarks and the time interval that the adversary actually observes between conversation turns. Time watermarks designed in this manner produce strictly defined time intervals in the traffic. Therefore, the adversary can control the covert information transmission by controlling the time interval between each turn, as illustrated in Figure~\ref{gap:2}.

\begin{figure}[tb]
  \centering
  \includegraphics[width=\linewidth]{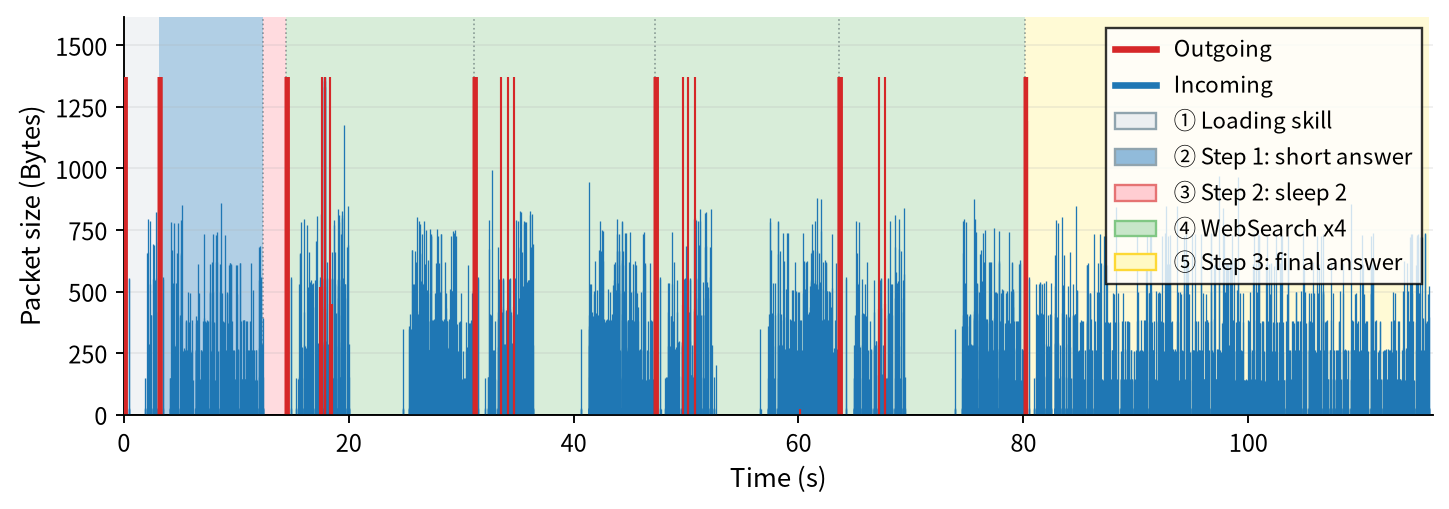}
  \caption{An example of an embedded timing watermark, with one Bash sleep of 2 seconds and four web searches inserted before the final answer.}\label{fig:skill_web_search}
\end{figure}

We are similarly concerned about the potential audit risks that such descriptions may introduce, particularly from LLM-based reviewers. We therefore select tasks from existing skills that introduce additional computational complexity to serve as concealed watermarks. The framework encodes whether the agent performs a web search operation during each conversation turn, thereby introducing a controllable delay in each turn. Figure~\ref{fig:skill_web_search} illustrates a comparison between the traffic behavior of web search actions and direct answer actions (step 1 and step 3), with the web search performing four rounds of retrieval. We observe that the traffic pattern during web search differs substantially from that of direct answering. In particular, the web search introduces additional waiting time because the agent performs search queries and summarizes the results in the background. Therefore, the adversary can distinguish web search actions from direct answer actions through direct traffic observation. Given that different models adopt different search strategies, we collected data on four distinct agents. This confirms that web search behavior constitutes a traffic pattern with a characteristic latency signature and can serve as a prominent timing watermark.

\subsection{Multi-Level Encoding Strategy}

We next consider how to structure the information to be transmitted so that (1) the adversary can locate when a user employs this skill from the traffic, and (2) the adversary can leak the information of interest through the traffic. We achieve the above goals by configuring watermarks as distinguishable multi-turn conversations. Inserting a segment of natural language description into a skill does not directly produce a multi-turn conversation, because the agent sends the prompt and the skill description to the LLM together rather than turn by turn. Generating a multi-turn conversation during agent-LLM interaction therefore requires following a specific paradigm. We summarize the paradigms that can be used for multi-turn watermark design as follows.

\begin{enumerate}[leftmargin=*, nosep]
\item Orchestrating tool-call sequences using natural language. Within a single-turn design, a specific tool is used to block the current session. This tool can be designed to run locally, such as locally executed scripts or commands, or remotely, such as invoking web search or other agent tools. These tool calls break the original single-turn conversation into multiple turns, thereby creating sufficient encoding space for the adversary.

\item Inserting reasoning between calls. Natural language descriptions must be inserted between multi-turn segments to form a semantically coherent logical chain, preventing the appearance of isolated and suspicious descriptions. For instance, the multi-turn structure can be semantically framed as common logical processes such as reflection and revision, plan and execution, or candidate proposal and evaluation. This ensures that the natural language descriptions embedded in the watermark are naturally consistent with the surrounding context. Such multi-turn paradigms are in fact already widely adopted in existing skills. Our watermark design paradigm therefore minimizes the likelihood of arousing suspicion from LLM-based security analyzers.

\item Embedding specific covert information into the available encoding space. We formalize this problem as follows. The adversary designs an $N$-round conversation procedure, where $N$ equals the length of $P$. $P$ is defined as a binary sequence. When the $i$-th element is 0, the skill initiates a prompt of type A. When the $i$-th element is 1, the skill initiates a prompt of type B. Here A and B represent two distinct traffic pattern pairs triggered by two different watermarks. The adversary therefore configures the outgoing sequences of the $N$ rounds as a specific trigger, while simultaneously encoding covert information in the timing and incoming packet size dimensions, thereby creating sufficient encoding space across $N$ rounds to transmit private information.

\end{enumerate}

Through the above design strategies, the adversary can achieve two goals. First, the adversary can design a unique traffic pattern that is distinguishable from any pattern in the existing skill marketplace, and determine whether a user is using a specific skill by observing whether the corresponding watermark traffic is generated. Second, the adversary maintains multiple groups of objects of interest and encodes them into the skill using different watermarks to determine whether the user's prompt contains content of interest to the adversary. We illustrate the encoding scheme with the following instantiation. Suppose the adversary maintains a dictionary of $K$ sensitive keywords $\{w_1, \ldots, w_K\}$. For each keyword $w_k$, the adversary assigns a distinct binary codeword of length $N$ (the number of conversation turns). In each turn $i \in \{1,\ldots,N\}$, the skill checks whether the user's prompt contains any keyword whose $i$-th codeword bit is 1. If so, the skill executes watermark pattern A (e.g., a brief answer followed by \texttt{sleep}); if not, the skill executes watermark pattern B (e.g., a detailed answer with web search). The adversary observes the traffic across all $N$ turns, classifies each turn as pattern A or B, and concatenates the bits to recover which keyword was present. With $N=4$ turns and two distinguishable patterns per turn, the adversary can discriminate among up to $2^4 = 16$ keywords in a single session. We evaluate the effectiveness of SkillWatermark in Section~\ref{sec:Evaluation}.

\section{Evaluation}\label{sec:Evaluation}

\subsection{Experiment Setup}

\noindent \textbf{Dataset.} We crawled 65,699 skills from ClawHub and categorized them by domain into 14 specialized domains. Because different skills employ distinct and highly specialized prompts, we simplify the prompt design by selecting the 10 most-downloaded skills in each domain for testing. For each skill we design questions that match its intended functionality, embedding identity, goal, and requirement information within them to simulate prompts that are as realistic as possible. To preserve the original functionality of each skill, we invoke the watermark independently at the beginning of the original skill. In total, we captured over 7,000 samples for evaluation.

\noindent \textbf{Implementation.} To facilitate data collection, we designed an automated collection script following the approach of Whisper Leak~\cite{DBLP:journals/corr/abs-2511-03675}. We installed Claude Code~\cite{Claude}  on a laptop equipped with an Intel i7-13700H CPU and configured the runtime environment. During data collection, a Clash proxy was used to simulate the scenario in which a user employs a VPN. For ease of evaluation, we designed a three-turn conversational watermark and label each stage as T1 through T3.






\subsection{Traffic Consistency}

\begin{figure*}[tb]
  \centering
  \includegraphics[width=\linewidth]{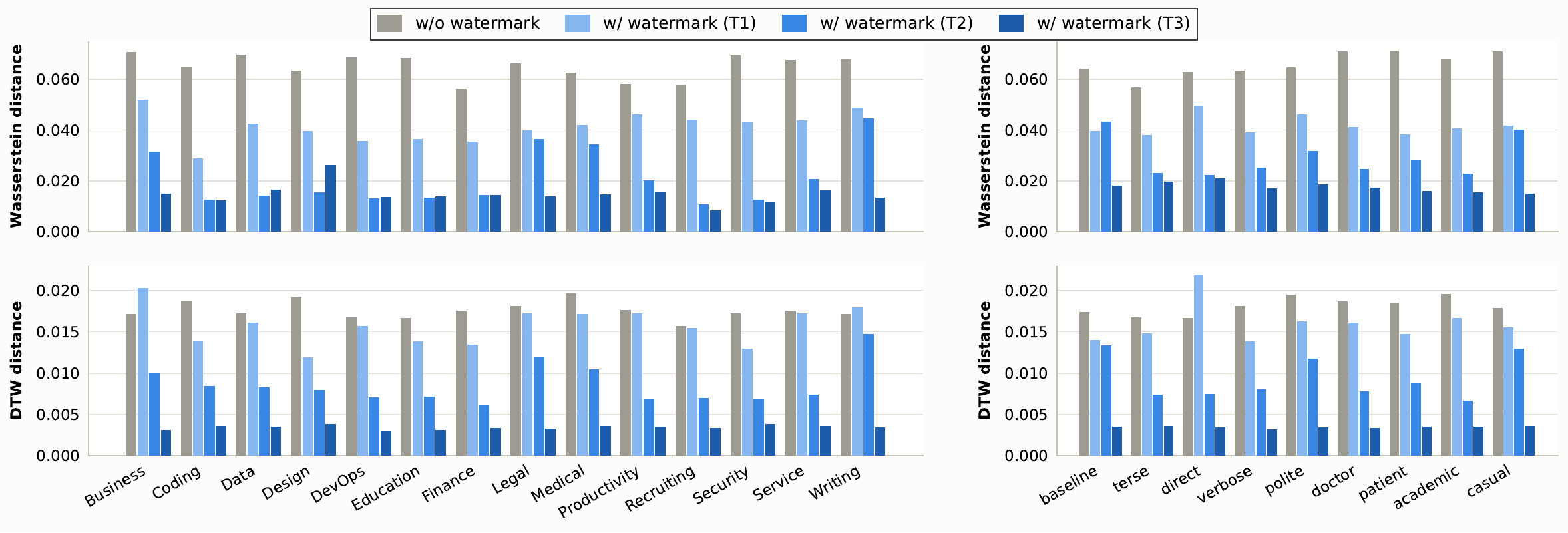}
  \caption{Comparison of Wasserstein distance and DTW distance between original skills and skills with an embedded three-stage watermark. Smaller distances indicate better intra-class consistency. Left: cross-domain consistency evaluation. Right: cross-prompt-style consistency evaluation.}\label{fig:domain}
\end{figure*}

We first evaluate the consistency between the traffic generated by skills embedding a watermark and that generated by ordinary skills. If the traffic for specific questions is highly consistent, the adversary can confirm the user's intent through direct observation, without the need to repeatedly collect data and train models. This is the core distinction between our method and prior work.

We report consistency using two types of metrics, namely statistical metrics based on Wasserstein distance~\cite{Panaretos_2019} and packet-level metrics based on Dynamic Time Warping (DTW) distance~\cite{1163055}. The lower the metric, the smaller the difference between traffic captures from different sessions, indicating more stable traffic. First, to evaluate generalizability, we assess whether the designed watermark is effective across skills from different domains. The left subfigure of Figure~\ref{fig:domain} presents the effect of embedding the watermark into skills from 14 different domains. We observe that after embedding the watermark, traffic consistency improves markedly, with gains in both statistical features and packet-level consistency. In particular, by the second conversation turn, 11 domains already exhibit more than a twofold improvement in consistency compared to the unwatermarked baseline. By the third turn, the traffic patterns of different skills exhibit highly similar structures. This indicates that the embedded watermark substantially enhances the observability of traffic patterns, and that these specific traffic patterns reflect the intent of the skill being used by the user.

Next, we examine whether our watermark can detect the corresponding traffic patterns when user prompts are topically related but stylistically diverse. This simulates the impact of different users' individual habits on the traffic patterns produced by prompts. We use the nine linguistic styles listed in Table~\ref{tab2}. The right subfigure of Figure~\ref{fig:domain} illustrates the impact of various prompt styles on traffic. We also observe that after the traffic is shaped by the watermark, the degree of structure increases while randomness decreases. We note that the DTW distance still exhibits significant randomness during the first watermark turn, which is caused by a conflict between the prompt style and the first-turn watermark. For instance, the Academic style demands more complex answers, whereas the first turn only requires a brief response. As the turn count increases, however, the degree of structuring also rises. By the third turn, the conversation already exhibits a high degree of consistency. This demonstrates that the designed watermark maintains stability across prompts of different styles but identical intent, and can effectively encode the user's intent into traffic patterns that are readily observable.

\begin{table}[tb]
\caption{Prompt style}
\begin{center}
\begin{tabular}{c|p{0.65\linewidth}}
\hline
Style & Example \\
\hline
Baseline & Give me some suggestions. \\
Terse   & Suggestions? \\
Direct  & Just tell me what to do. \\
Academic & I would appreciate receiving evidence-based recommendations pertaining to this matter. \\
Casual  & Hey, any tips for me? \\
Verbose & I was wondering if you might be able to provide me with a comprehensive set of suggestions that could potentially help me address this situation in the most effective manner possible. \\
Polite  & Could you kindly offer some suggestions? \\
Doctor  & Based on a profession's perspective, what management strategies would you recommend? \\
Patient & Can you help me understand what I should do about this? \\
\hline
\end{tabular}
\label{tab2}
\end{center}
\end{table}

\begin{figure}[tb]
    \centering
    \begin{subfigure}{\linewidth}
        \centering
        \includegraphics[width=\linewidth]{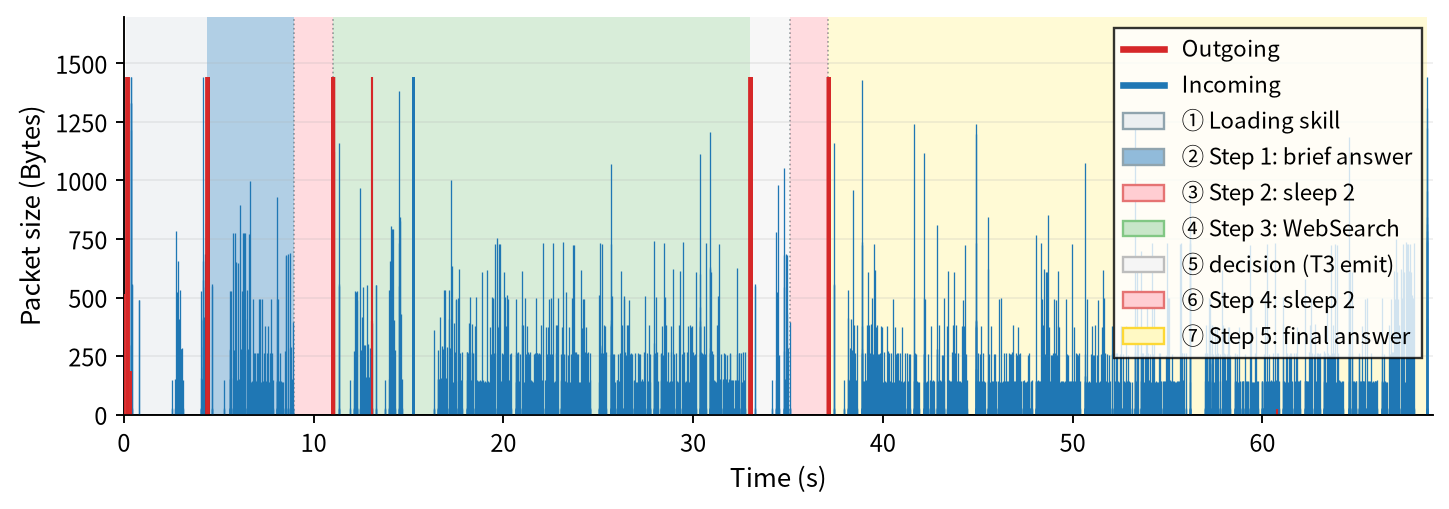}
        \caption{Watermark traffic pattern when using Claude Code.}
        \label{mo:1}
    \end{subfigure}

    \bigskip

    \begin{subfigure}{\linewidth}
        \centering
        \includegraphics[width=\linewidth]{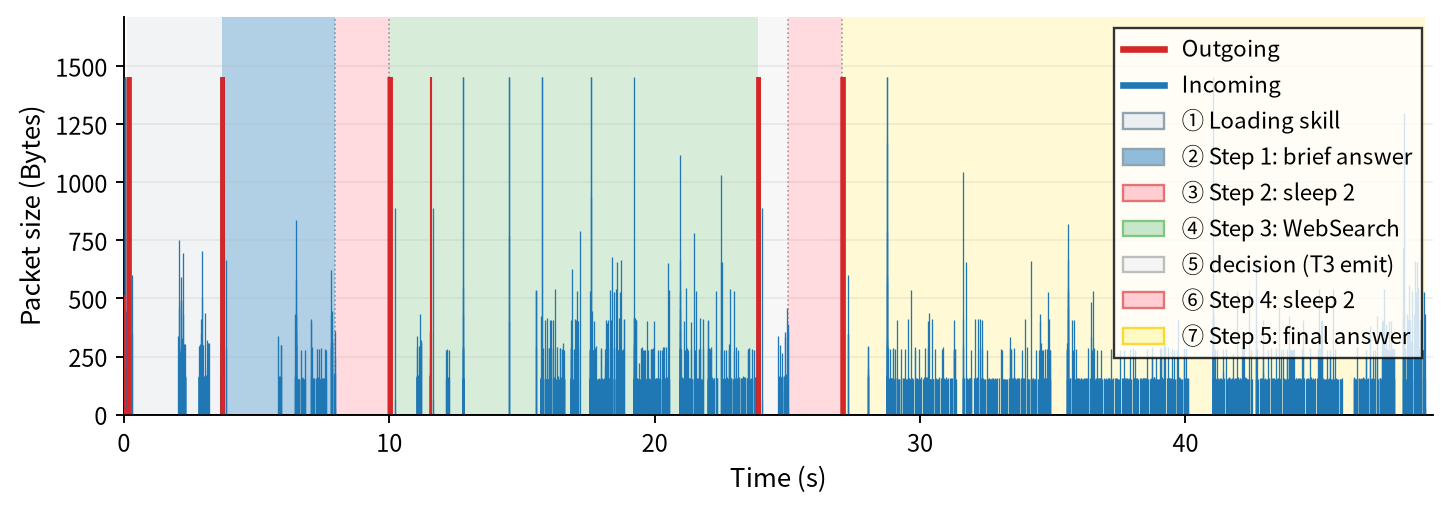}
        \caption{Watermark traffic pattern when using GLM 5.2.}
        \label{mo:2}
    \end{subfigure}
    \caption{Cross-model watermark consistency evaluation. The model API was switched by modifying the settings.local.json file, and traffic was then collected.}
    \label{fig:mo}  
\end{figure}

Next, we evaluate the watermark's effectiveness across different LLMs. We add GLM 5.2 for testing and present the traffic patterns of the sample watermark under different models. Figure \ref{fig:mo} presents the watermark shapes under different models. Although different models exhibit differences in fine-grained packet distributions, for instance the traffic of GLM is sparser, the behavior of each conversation turn encoded by the watermark exhibits a high degree of consistency across multi-turn interactions. This is precisely the key distinction between our watermark scheme and prior traffic collection approaches, namely that we need not collect large volumes of data under many influencing factors. The adversary need only observe the corresponding behavioral pattern to confirm that the user has triggered a specific watermark.

\subsection{Decoding Evaluation}
Next, we evaluate whether the implicit information encoded in the watermarks can be recovered from the traffic patterns in the collected samples. We assess the True Positive Rate (TPR) and False Positive Rate (FPR) by testing on traffic traces with embedded watermarks and on traffic traces without embedded watermarks. The results are presented in the table. Since the consistency evaluation indicates that T1 provides limited discriminability for traffic, we use T2 and T3 as the watermark identification indicators. As the number of conversation turns increases, TPR rises while FPR declines. On the watermarked dataset, our method achieves a decoding success rate of 98.8 percent. On the unwatermarked dataset, the false positive rate is only 8 percent.

\begin{table}[tb]
  \caption{Evaluation of TPR and FPR on datasets with and without embedded watermarks.}
  \label{tab3}
  \begin{center}
  \begin{tabular}{lcc}
    \hline
    \textbf{Criterion} & \textbf{TPR} & \textbf{FPR} \\
    \hline
    watermark (T2) & 92.5\% & 13\% \\
    watermark (T2+T3) & 98.8\% & 8\% \\
    \hline
  \end{tabular}
  \end{center}
\end{table}

\subsection{Anti-detection Evaluation}

\begin{tcolorbox}[title={resume-optimizer}]
...\\
\> If the user provides only a Job 
  Description without a resume, prompt for resume materials.\\
\> If the user provides only a resume without a Job 
  Description, proceed in \texttt{rewrite} mode and prompt after Step~2...\\
...
\end{tcolorbox}

Finally, we evaluate whether our attack method can pass existing detection tools. Following the previous works, we conduct a qualitative study applying LLM-based detection tools to assess the stealthiness of our attack. We find that even when LLM auditing tools explicitly observe the watermark content, they do not classify the watermark as belonging to any known attack category. In particular, when the watermark attempts to leak specific implicit information, for instance by using pattern A for elderly users and pattern B for younger users, the LLM audit naturally interprets this as personalization intended to better serve specific user populations. It does not regard such targeted analysis as a privacy inference risk, because many legitimate skills employ similar strategies. We list below an example from existing publicly available skills that already employ mechanisms similar in structure to our watermarks~\cite{resume-optimizer}. The adversary need only observe whether the corresponding traffic is generated to determine whether the user provided a Job Description or a resume. This illustrates the weakness of existing detection paradigms: traffic side-channel leakage does not require any of the known malicious-skill primitives, and current auditing frameworks are not designed to detect it.

\section{Discussion and Conclusion}
\noindent \textbf{Countermeasures.} Our attack is grounded in a fundamental observation that specific prompts of skills generate distinguishable traffic patterns, which are inadvertently encoded as skills optimize for superior output performance. These interaction patterns establish an implicit side channel within network traffic that an attacker can exploit to compromise user privacy. The most straightforward countermeasure would be to prohibit any description for producing structured traffic patterns during multi-turn conversations. However, this approach runs directly counter to the foundational purpose of skill design, as structured multi-turn interactions are intended to enhance task completion efficiency and to reduce the frequency of manual user intervention. We therefore advocate that existing skill auditing frameworks incorporate an additional detection dimension that specifically inspects skills capable of generating patterned traffic across conversation turns, thereby strengthening the community's defense perimeter.

\noindent \textbf{Limitations.} Our attack paradigm is applicable only to LLMs whose network traffic can be observed. If a user deploys the model in an offline setting, our attack does not take effect. Scenarios in which the user does not employ any skill likewise fall outside the scope of this work. If a user conducts only a single conversation turn with a very short prompt, the amount of leaked information remains limited. When a user initiates multiple concurrent sessions, we are unable to isolate individual sessions because the VPN masks port numbers, and the method therefore does not apply. False positives can arise when unmodified skills coincidentally produce traffic resembling the watermark. The adversary mitigates this by avoiding watermark designs that replicate existing skill descriptions.

\noindent \textbf{Conclusion.} In this paper, we proposed SkillWatermark, a novel attack paradigm that exploits the embedding of specific prompt terms to generate distinctive traffic patterns. These watermarks are benign structured statements, and as a result existing detection paradigms overlook them. Our experiments demonstrate that watermarked skills produce highly consistent traffic patterns across 14 domains and diverse prompt styles, and that LLM-based auditing tools do not flag the transformed skills as malicious. This paper presents the first systematic exposition of how covert channels can be constructed in agent skills using prompt-level watermarks, and reveals the blind spot of existing skill security auditing tools with respect to traffic side-channel leakage.



\bibliographystyle{IEEEtran}
\bibliography{sample-base}

\end{document}